\documentstyle[11pt,a4,sw20lart]{article}

\makeatletter
\newcount\@hour\newcount\@minute\chardef\@x10\chardef\@xv60
\def\tcitime{
\def\@time{%
  \@minute\time\@hour\@minute\divide\@hour\@xv
  \ifnum\@hour<\@x 0\fi\the\@hour:%
  \multiply\@hour\@xv\advance\@minute-\@hour
  \ifnum\@minute<\@x 0\fi\the\@minute
  }}%

\@ifundefined{hyperref}{}{}

\@ifundefined{qExtProgCall}{\def\qExtProgCall#1#2#3#4#5#6{\relax}}{}
\def\QCTOpt[#1]#2{%
  \def\QCTOptB{#1}
  \def\QCTOptA{#2}
}
\def\QCTNOpt#1{%
  \def\QCTOptA{#1}
  \let\QCTOptB\empty
}
\def\Qct{%
  \@ifnextchar[{%
    \QCTOpt}{\QCTNOpt}
}
\def\QCBOpt[#1]#2{%
  \def\QCBOptB{#1}
  \def\QCBOptA{#2}
}
\def\QCBNOpt#1{%
  \def\QCBOptA{#1}
  \let\QCBOptB\empty
}
\def\Qcb{%
  \@ifnextchar[{%
    \QCBOpt}{\QCBNOpt}
}
\def\PrepCapArgs{%
  \ifx\QCBOptA\empty
    \ifx\QCTOptA\empty
      {}%
    \else
      \ifx\QCTOptB\empty
        {\QCTOptA}%
      \else
        [\QCTOptB]{\QCTOptA}%
      \fi
    \fi
  \else
    \ifx\QCBOptA\empty
      {}%
    \else
      \ifx\QCBOptB\empty
        {\QCBOptA}%
      \else
        [\QCBOptB]{\QCBOptA}%
      \fi
    \fi
  \fi
}
\newcount\GRAPHICSTYPE
\GRAPHICSTYPE=\z@
\def\GRAPHICSPS#1{%
 \ifcase\GRAPHICSTYPE
   \special{ps: #1}%
 \or
   \special{language "PS", include "#1"}%
 \fi
}%
\def\graffile#1#2#3#4{%
    \leavevmode
    \raise -#4 \BOXTHEFRAME{%
        \hbox to #2{\raise #3\hbox to #2{\null #1\hfil}}}%
}%
\def\draftbox#1#2#3#4{%
 \leavevmode\raise -#4 \hbox{%
  \frame{\rlap{\protect\tiny #1}\hbox to #2%
   {\vrule height#3 width\z@ depth\z@\hfil}%
  }%
 }%
}%
\newcount\draft
\draft=\z@

\newif\ifwasdraft
\wasdraftfalse

\def\GRAPHIC#1#2#3#4#5{%
 \ifnum\draft=\@ne\draftbox{#2}{#3}{#4}{#5}%
  \else\graffile{#1}{#3}{#4}{#5}%
  \fi
 }%
\def\addtoLaTeXparams#1{%
    \edef\LaTeXparams{\LaTeXparams #1}}%
\newif\ifBoxFrame \BoxFramefalse
\newif\ifOverFrame \OverFramefalse
\newif\ifUnderFrame \UnderFramefalse

\def\BOXTHEFRAME#1{%
   \hbox{%
      \ifBoxFrame
         \frame{#1}%
      \else
         {#1}%
      \fi
   }%
}

\def\doFRAMEparams#1{\BoxFramefalse\OverFramefalse\UnderFramefalse\readFRAMEparams#1\end}%
\def\readFRAMEparams#1{%
 \ifx#1\end%
  \let\next=\relax
  \else
  \ifx#1i\dispkind=\z@\fi
  \ifx#1d\dispkind=\@ne\fi
  \ifx#1f\dispkind=\tw@\fi
  \ifx#1t\addtoLaTeXparams{t}\fi
  \ifx#1b\addtoLaTeXparams{b}\fi
  \ifx#1p\addtoLaTeXparams{p}\fi
  \ifx#1h\addtoLaTeXparams{h}\fi
  \ifx#1X\BoxFrametrue\fi
  \ifx#1O\OverFrametrue\fi
  \ifx#1U\UnderFrametrue\fi
  \ifx#1w
    \ifnum\draft=1\wasdrafttrue\else\wasdraftfalse\fi
    \draft=\@ne
  \fi
  \let\next=\readFRAMEparams
  \fi
 \next
 }%
\def\IFRAME#1#2#3#4#5#6{%
      \bgroup
      \let\QCTOptA\empty
      \let\QCTOptB\empty
      \let\QCBOptA\empty
      \let\QCBOptB\empty
      #6%
      \parindent=0pt%
      \leftskip=0pt
      \rightskip=0pt
      \setbox0 = \hbox{\QCBOptA}%
      \@tempdima = #1\relax
      \ifOverFrame
          \typeout{This is not implemented yet}%
          \show\HELP
      \else
         \ifdim\wd0>\@tempdima
            \advance\@tempdima by \@tempdima
            \ifdim\wd0 >\@tempdima
               \textwidth=\@tempdima
               \setbox1 =\vbox{%
                  \noindent\hbox to \@tempdima{\hfill\GRAPHIC{#5}{#4}{#1}{#2}{#3}\hfill}\\%
                  \noindent\hbox to \@tempdima{\parbox[b]{\@tempdima}{\QCBOptA}}%
               }%
               \wd1=\@tempdima
            \else
               \textwidth=\wd0
               \setbox1 =\vbox{%
                 \noindent\hbox to \wd0{\hfill\GRAPHIC{#5}{#4}{#1}{#2}{#3}\hfill}\\%
                 \noindent\hbox{\QCBOptA}%
               }%
               \wd1=\wd0
            \fi
         \else
            \ifdim\wd0>0pt
              \hsize=\@tempdima
              \setbox1 =\vbox{%
                \unskip\GRAPHIC{#5}{#4}{#1}{#2}{0pt}%
                \break
                \unskip\hbox to \@tempdima{\hfill \QCBOptA\hfill}%
              }%
              \wd1=\@tempdima
           \else
              \hsize=\@tempdima
              \setbox1 =\vbox{%
                \unskip\GRAPHIC{#5}{#4}{#1}{#2}{0pt}%
              }%
              \wd1=\@tempdima
           \fi
         \fi
         \@tempdimb=\ht1
         \advance\@tempdimb by \dp1
         \advance\@tempdimb by -#2%
         \advance\@tempdimb by #3%
         \leavevmode
         \raise -\@tempdimb \hbox{\box1}%
      \fi
      \egroup%
}%
\def\DFRAME#1#2#3#4#5{%
 \begin{center}
     \let\QCTOptA\empty
     \let\QCTOptB\empty
     \let\QCBOptA\empty
     \let\QCBOptB\empty
     \ifOverFrame 
        #5\QCTOptA\par
     \fi
     \GRAPHIC{#4}{#3}{#1}{#2}{\z@}
     \ifUnderFrame 
        \nobreak\par #5\QCBOptA
     \fi
 \end{center}%
 }%
\def\FFRAME#1#2#3#4#5#6#7{%
 \begin{figure}[#1]%
  \let\QCTOptA\empty
  \let\QCTOptB\empty
  \let\QCBOptA\empty
  \let\QCBOptB\empty
  \ifOverFrame
    #4
    \ifx\QCTOptA\empty
    \else
      \ifx\QCTOptB\empty
        \caption{\QCTOptA}%
      \else
        \caption[\QCTOptB]{\QCTOptA}%
      \fi
    \fi
    \ifUnderFrame\else
      \label{#5}%
    \fi
  \else
    \UnderFrametrue%
  \fi
  \begin{center}\GRAPHIC{#7}{#6}{#2}{#3}{\z@}\end{center}%
  \ifUnderFrame
    #4
    \ifx\QCBOptA\empty
      \caption{}%
    \else
      \ifx\QCBOptB\empty
        \caption{\QCBOptA}%
      \else
        \caption[\QCBOptB]{\QCBOptA}%
      \fi
    \fi
    \label{#5}%
  \fi
  \end{figure}%
 }%
\newcount\dispkind%

\def\makeactives{
  \catcode`\"=\active
  \catcode`\;=\active
  \catcode`\:=\active
  \catcode`\'=\active
  \catcode`\~=\active
}
\bgroup
   \makeactives
   \gdef\activesoff{%
      \def"{\string"}
      \def;{\string;}
      \def:{\string:}
      \def'{\string'}
      \def~{\string~}
    }
\egroup

\def\FRAME#1#2#3#4#5#6#7#8{%
 \bgroup
 \@ifundefined{bbl@deactivate}{}{\activesoff}
 \ifnum\draft=\@ne
   \wasdrafttrue
 \else
   \wasdraftfalse%
 \fi
 \def\LaTeXparams{}%
 \dispkind=\z@
 \def\LaTeXparams{}%
 \doFRAMEparams{#1}%
 \ifnum\dispkind=\z@\IFRAME{#2}{#3}{#4}{#7}{#8}{#5}\else
  \ifnum\dispkind=\@ne\DFRAME{#2}{#3}{#7}{#8}{#5}\else
   \ifnum\dispkind=\tw@
    \edef\@tempa{\noexpand\FFRAME{\LaTeXparams}}%
    \@tempa{#2}{#3}{#5}{#6}{#7}{#8}%
    \fi
   \fi
  \fi
  \ifwasdraft\draft=1\else\draft=0\fi{}%
  \egroup
 }%
\def\TEXUX#1{"texux"}

\def\func#1{\mathop{\rm #1}}%
\long\def\QQQ#1#2{%
     \long\expandafter\def\csname#1\endcsname{#2}}%
\@ifundefined{QTP}{\def\QTP#1{}}{}
\@ifundefined{QEXCLUDE}{\def\QEXCLUDE#1{}}{}
\@ifundefined{Qlb}{}{}
\@ifundefined{Qlt}{}{}
\long\def\QQA#1#2{}%
\def\QTR#1#2{{\csname#1\endcsname #2}}
\def\EXPAND#1[#2]#3{}%
\def\NOEXPAND#1[#2]#3{}%
\def\LaTeXparent#1{}%
\def\ChildStyles#1{}%
\def\ChildDefaults#1{}%
\def\QTagDef#1#2#3{}%
\@ifundefined{StyleEditBeginDoc}{}{}
\def\QQfnmark#1{\footnotemark}

\@ifundefined{INDEX}{\def\INDEX#1#2{}{}}{}%
\@ifundefined{SUBINDEX}{\def\SUBINDEX#1#2#3{}{}{}}{}%
\@ifundefined{initial}%
   {\def\initial#1{\bigbreak{\raggedright\large\bf #1}\kern 2\p@\penalty3000}}%
   {}%
\@ifundefined{entry}{}{}%
\@ifundefined{primary}{}{}%
\@ifundefined{secondary}{}{}%
\@ifundefined{ZZZ}{}{\makeatletter\input gnuindex.sty\makeatother\makeindex\makeatletter}%
\@ifundefined{abstract}{%
 \def\abstract{%
  \if@twocolumn
   \section*{Abstract (Not appropriate in this style!)}%
   \else \small 
   \begin{center}{\bf Abstract\vspace{-.5em}\vspace{\z@}}\end{center}%
   \quotation 
   \fi
  }%
 }{%
 }%
\@ifundefined{endabstract}{\def\endabstract
  {\if@twocolumn\else\endquotation\fi}}{}%
\@ifundefined{maketitle}{\def\maketitle#1{}}{}%
\@ifundefined{affiliation}{\def\affiliation#1{}}{}%
\@ifundefined{proof}{}{}%
\@ifundefined{endproof}{}{}%
\@ifundefined{newfield}{\def\newfield#1#2{}}{}%
\@ifundefined{chapter}{\def\chapter#1{\par(Chapter head:)#1\par }%
 \newcount\c@chapter}{}%
\@ifundefined{part}{\def\part#1{\par(Part head:)#1\par }}{}%
\@ifundefined{section}{\def\section#1{\par(Section head:)#1\par }}{}%
\@ifundefined{subsection}{\def\subsection#1%
 {\par(Subsection head:)#1\par }}{}%
\@ifundefined{subsubsection}{\def\subsubsection#1%
 {\par(Subsubsection head:)#1\par }}{}%
\@ifundefined{paragraph}{\def\paragraph#1%
 {\par(Subsubsubsection head:)#1\par }}{}%
\@ifundefined{subparagraph}{\def\subparagraph#1%
 {\par(Subsubsubsubsection head:)#1\par }}{}%
\@ifundefined{therefore}{}{}%
\@ifundefined{backepsilon}{}{}%
\@ifundefined{yen}{}{}%
\@ifundefined{registered}{%
   \def\registered{\relax\ifmmode{}\r@gistered
                    \else$\m@th\r@gistered$\fi}%
 \def\r@gistered{^{\ooalign
  {\hfil\raise.07ex\hbox{$\scriptstyle\rm\text{R}$}\hfil\crcr
  \mathhexbox20D}}}}{}%
\@ifundefined{Eth}{}{}%
\@ifundefined{eth}{}{}%
\@ifundefined{Thorn}{}{}%
\@ifundefined{thorn}{}{}%
\@ifundefined{degree}{}{}%
\newdimen\theight
\def\Column{%
 \vadjust{\setbox\z@=\hbox{\scriptsize\quad\quad tcol}%
  \theight=\ht\z@\advance\theight by \dp\z@\advance\theight by \lineskip
  \kern -\theight \vbox to \theight{%
   \rightline{\rlap{\box\z@}}%
   \vss
   }%
  }%
 }%
\def\qed{%
 \ifhmode\unskip\nobreak\fi\ifmmode\ifinner\else\hskip5\p@\fi\fi
 \hbox{\hskip5\p@\vrule width4\p@ height6\p@ depth1.5\p@\hskip\p@}%
 }%
\def\miss{\hbox{\vrule height2\p@ width 2\p@ depth\z@}}%
\def\tcol#1{{\baselineskip=6\p@ \vcenter{#1}} \Column}  %
\def\newfmtname{LaTeX2e}
\def\chkcompat{%
   \if@compatibility
   \else
     \usepackage{latexsym}
   \fi
}

\ifx\fmtname\newfmtname
  \DeclareOldFontCommand{\rm}{\normalfont\rmfamily}{\mathrm}
  \DeclareOldFontCommand{\sf}{\normalfont\sffamily}{\mathsf}
  \DeclareOldFontCommand{\tt}{\normalfont\ttfamily}{\mathtt}
  \DeclareOldFontCommand{\bf}{\normalfont\bfseries}{\mathbf}
  \DeclareOldFontCommand{\it}{\normalfont\itshape}{\mathit}
  \DeclareOldFontCommand{\sl}{\normalfont\slshape}{\@nomath\sl}
  \DeclareOldFontCommand{\sc}{\normalfont\scshape}{\@nomath\sc}
  \chkcompat
\fi

\def\alpha{{\Greekmath 010B}}%
\def\beta{{\Greekmath 010C}}%
\def\gamma{{\Greekmath 010D}}%
\def\delta{{\Greekmath 010E}}%
\def\epsilon{{\Greekmath 010F}}%
\def\zeta{{\Greekmath 0110}}%
\def\eta{{\Greekmath 0111}}%
\def\theta{{\Greekmath 0112}}%
\def\iota{{\Greekmath 0113}}%
\def\kappa{{\Greekmath 0114}}%
\def\lambda{{\Greekmath 0115}}%
\def\mu{{\Greekmath 0116}}%
\def\nu{{\Greekmath 0117}}%
\def\xi{{\Greekmath 0118}}%
\def\pi{{\Greekmath 0119}}%
\def\rho{{\Greekmath 011A}}%
\def\sigma{{\Greekmath 011B}}%
\def\tau{{\Greekmath 011C}}%
\def\upsilon{{\Greekmath 011D}}%
\def\phi{{\Greekmath 011E}}%
\def\chi{{\Greekmath 011F}}%
\def\psi{{\Greekmath 0120}}%
\def\omega{{\Greekmath 0121}}%
\def\varepsilon{{\Greekmath 0122}}%
\def\vartheta{{\Greekmath 0123}}%
\def\varpi{{\Greekmath 0124}}%
\def\varrho{{\Greekmath 0125}}%
\def\varsigma{{\Greekmath 0126}}%
\def\varphi{{\Greekmath 0127}}%

\def\nabla{{\Greekmath 0272}}
\def\FindBoldGroup{%
   {\setbox0=\hbox{$\mathbf{x\global\edef\theboldgroup{\the\mathgroup}}$}}%
}

\def\Greekmath#1#2#3#4{%
    \if@compatibility
        \ifnum\mathgroup=\symbold
           \mathchoice{\mbox{\boldmath$\displaystyle\mathchar"#1#2#3#4$}}%
                      {\mbox{\boldmath$\textstyle\mathchar"#1#2#3#4$}}%
                      {\mbox{\boldmath$\scriptstyle\mathchar"#1#2#3#4$}}%
                      {\mbox{\boldmath$\scriptscriptstyle\mathchar"#1#2#3#4$}}%
        \else
           \mathchar"#1#2#3#4%
        \fi 
    \else 
        \FindBoldGroup
        \ifnum\mathgroup=\theboldgroup 
           \mathchoice{\mbox{\boldmath$\displaystyle\mathchar"#1#2#3#4$}}%
                      {\mbox{\boldmath$\textstyle\mathchar"#1#2#3#4$}}%
                      {\mbox{\boldmath$\scriptstyle\mathchar"#1#2#3#4$}}%
                      {\mbox{\boldmath$\scriptscriptstyle\mathchar"#1#2#3#4$}}%
        \else
           \mathchar"#1#2#3#4%
        \fi     	    
	  \fi}

\newif\ifGreekBold  \GreekBoldfalse
\let\SAVEPBF=\pbf
\def\pbf{\GreekBoldtrue\SAVEPBF}%

\@ifundefined{theorem}{}{}
\@ifundefined{lemma}{}{}
\@ifundefined{corollary}{}{}
\@ifundefined{conjecture}{}{}
\@ifundefined{proposition}{}{}
\@ifundefined{axiom}{}{}
\@ifundefined{remark}{}{}
\@ifundefined{example}{}{}
\@ifundefined{exercise}{}{}
\@ifundefined{definition}{}{}

\@ifundefined{mathletters}{%
  \newcounter{equationnumber}  
  \def\mathletters{%
     \addtocounter{equation}{1}
     \edef\@currentlabel{\theequation}%
     \setcounter{equationnumber}{\c@equation}
     \setcounter{equation}{0}%
     \edef\theequation{\@currentlabel\noexpand\alph{equation}}%
  }
  
}{}

\@ifundefined{BibTeX}{%
    \def\BibTeX{{\rm B\kern-.05em{\sc i\kern-.025em b}\kern-.08em
                 T\kern-.1667em\lower.7ex\hbox{E}\kern-.125emX}}}{}%
\@ifundefined{AmS}%
    {\def\AmS{{\protect\usefont{OMS}{cmsy}{m}{n}%
                A\kern-.1667em\lower.5ex\hbox{M}\kern-.125emS}}}{}%
\@ifundefined{AmSTeX}{}{}%
\ifx\ds@amstex\relax
\makeatother\endinput
\else
   \@ifpackageloaded{amstex}%
      {\message{amstex already loaded}\makeatother\endinput}
      {}
   \@ifpackageloaded{amsgen}%
      {\message{amsgen already loaded}\makeatother\endinput}
      {}
\fi
\let\DOTSI\relax
\def\RIfM@{\relax\ifmmode}%
\def\FN@{\futurelet\next}%
\newcount\intno@
\def\iint{\DOTSI\intno@\tw@\FN@\ints@}%
\def\iiint{\DOTSI\intno@\thr@@\FN@\ints@}%
\def\iiiint{\DOTSI\intno@4 \FN@\ints@}%
\def\idotsint{\DOTSI\intno@\z@\FN@\ints@}%
\def\ints@{\findlimits@\ints@@}%
\newif\iflimtoken@
\newif\iflimits@
\def\findlimits@{\limtoken@true\ifx\next\limits\limits@true
 \else\ifx\next\nolimits\limits@false\else
 \limtoken@false\ifx\ilimits@\nolimits\limits@false\else
 \ifinner\limits@false\else\limits@true\fi\fi\fi\fi}%
\def\multint@{\int\ifnum\intno@=\z@\intdots@                          
 \else\intkern@\fi                                                    
 \ifnum\intno@>\tw@\int\intkern@\fi                                   
 \ifnum\intno@>\thr@@\int\intkern@\fi                                 
 \int}
\def\multintlimits@{\intop\ifnum\intno@=\z@\intdots@\else\intkern@\fi
 \ifnum\intno@>\tw@\intop\intkern@\fi
 \ifnum\intno@>\thr@@\intop\intkern@\fi\intop}%
\def\intic@{%
    \mathchoice{\hskip.5em}{\hskip.4em}{\hskip.4em}{\hskip.4em}}%
\def\negintic@{\mathchoice
 {\hskip-.5em}{\hskip-.4em}{\hskip-.4em}{\hskip-.4em}}%
\def\ints@@{\iflimtoken@                                              
 \def\ints@@@{\iflimits@\negintic@
   \mathop{\intic@\multintlimits@}\limits                             
  \else\multint@\nolimits\fi                                          
  \eat@}
 \else                                                                
 \def\ints@@@{\iflimits@\negintic@
  \mathop{\intic@\multintlimits@}\limits\else
  \multint@\nolimits\fi}\fi\ints@@@}%
\def\intkern@{\mathchoice{\!\!\!}{\!\!}{\!\!}{\!\!}}%
\def\plaincdots@{\mathinner{\cdotp\cdotp\cdotp}}%
\def\intdots@{\mathchoice{\plaincdots@}%
 {{\cdotp}\mkern1.5mu{\cdotp}\mkern1.5mu{\cdotp}}%
 {{\cdotp}\mkern1mu{\cdotp}\mkern1mu{\cdotp}}%
 {{\cdotp}\mkern1mu{\cdotp}\mkern1mu{\cdotp}}}%
\def\RIfM@{\relax\protect\ifmmode}
\def\text{\RIfM@\expandafter\text@\else\expandafter\mbox\fi}
\let\nfss@text\text
\def\text@#1{\mathchoice
   {\textdef@\displaystyle\f@size{#1}}%
   {\textdef@\textstyle\tf@size{\firstchoice@false #1}}%
   {\textdef@\textstyle\sf@size{\firstchoice@false #1}}%
   {\textdef@\textstyle \ssf@size{\firstchoice@false #1}}%
   \glb@settings}

\def\textdef@#1#2#3{\hbox{{%
                    \everymath{#1}%
                    \let\f@size#2\selectfont
                    #3}}}
\newif\iffirstchoice@
\firstchoice@true
\def\Let@{\relax\iffalse{\fi\let\\=\cr\iffalse}\fi}%
\def\vspace@{\def\vspace##1{\crcr\noalign{\vskip##1\relax}}}%
\def\multilimits@{\bgroup\vspace@\Let@
 \baselineskip\fontdimen10 \scriptfont\tw@
 \advance\baselineskip\fontdimen12 \scriptfont\tw@
 \lineskip\thr@@\fontdimen8 \scriptfont\thr@@
 \lineskiplimit\lineskip
 \vbox\bgroup\ialign\bgroup\hfil$\m@th\scriptstyle{##}$\hfil\crcr}%
\def\Sb{_\multilimits@}%
\def\endSb{\crcr\egroup\egroup\egroup}%
\def\Sp{^\multilimits@}%

\newdimen\ex@
\def\rightarrowfill@#1{$#1\m@th\mathord-\mkern-6mu\cleaders
 \hbox{$#1\mkern-2mu\mathord-\mkern-2mu$}\hfill
 \mkern-6mu\mathord\rightarrow$}%
\def\leftarrowfill@#1{$#1\m@th\mathord\leftarrow\mkern-6mu\cleaders
 \hbox{$#1\mkern-2mu\mathord-\mkern-2mu$}\hfill\mkern-6mu\mathord-$}%
\def\leftrightarrowfill@#1{$#1\m@th\mathord\leftarrow
\mkern-6mu\cleaders
 \hbox{$#1\mkern-2mu\mathord-\mkern-2mu$}\hfill
 \mkern-6mu\mathord\rightarrow$}%
\def\overrightarrow{\mathpalette\overrightarrow@}%
\def\overrightarrow@#1#2{\vbox{\ialign{##\crcr\rightarrowfill@#1\crcr
 \noalign{\kern-\ex@\nointerlineskip}$\m@th\hfil#1#2\hfil$\crcr}}}%

\def\overleftarrow{\mathpalette\overleftarrow@}%
\def\overleftarrow@#1#2{\vbox{\ialign{##\crcr\leftarrowfill@#1\crcr
 \noalign{\kern-\ex@\nointerlineskip}$\m@th\hfil#1#2\hfil$\crcr}}}%
\def\overleftrightarrow{\mathpalette\overleftrightarrow@}%
\def\overleftrightarrow@#1#2{\vbox{\ialign{##\crcr
   \leftrightarrowfill@#1\crcr
 \noalign{\kern-\ex@\nointerlineskip}$\m@th\hfil#1#2\hfil$\crcr}}}%
\def\underrightarrow{\mathpalette\underrightarrow@}%
\def\underrightarrow@#1#2{\vtop{\ialign{##\crcr$\m@th\hfil#1#2\hfil
  $\crcr\noalign{\nointerlineskip}\rightarrowfill@#1\crcr}}}%

\def\underleftarrow{\mathpalette\underleftarrow@}%
\def\underleftarrow@#1#2{\vtop{\ialign{##\crcr$\m@th\hfil#1#2\hfil
  $\crcr\noalign{\nointerlineskip}\leftarrowfill@#1\crcr}}}%
\def\underleftrightarrow{\mathpalette\underleftrightarrow@}%
\def\underleftrightarrow@#1#2{\vtop{\ialign{##\crcr$\m@th
  \hfil#1#2\hfil$\crcr
 \noalign{\nointerlineskip}\leftrightarrowfill@#1\crcr}}}%

\def\qopnamewl@#1{\mathop{\operator@font#1}\nlimits@}
\let\nlimits@\displaylimits
\def\setboxz@h{\setbox\z@\hbox}

\def\varlim@#1#2{\mathop{\vtop{\ialign{##\crcr
 \hfil$#1\m@th\operator@font lim$\hfil\crcr
 \noalign{\nointerlineskip}#2#1\crcr
 \noalign{\nointerlineskip\kern-\ex@}\crcr}}}}

 \def\rightarrowfill@#1{\m@th\setboxz@h{$#1-$}\ht\z@\z@
  $#1\copy\z@\mkern-6mu\cleaders
  \hbox{$#1\mkern-2mu\box\z@\mkern-2mu$}\hfill
  \mkern-6mu\mathord\rightarrow$}
\def\leftarrowfill@#1{\m@th\setboxz@h{$#1-$}\ht\z@\z@
  $#1\mathord\leftarrow\mkern-6mu\cleaders
  \hbox{$#1\mkern-2mu\copy\z@\mkern-2mu$}\hfill
  \mkern-6mu\box\z@$}

\def\projlim{\qopnamewl@{proj\,lim}}
\def\injlim{\qopnamewl@{inj\,lim}}
\def\varinjlim{\mathpalette\varlim@\rightarrowfill@}
\def\varprojlim{\mathpalette\varlim@\leftarrowfill@}
\def\varliminf{\mathpalette\varliminf@{}}
\def\varliminf@#1{\mathop{\underline{\vrule\@depth.2\ex@\@width\z@
   \hbox{$#1\m@th\operator@font lim$}}}}
\def\varlimsup{\mathpalette\varlimsup@{}}
\def\varlimsup@#1{\mathop{\overline
  {\hbox{$#1\m@th\operator@font lim$}}}}

\def\stackunder#1#2{\mathrel{\mathop{#2}\limits_{#1}}}%
\begingroup \catcode `|=0 \catcode `[= 1
\catcode`]=2 \catcode `\{=12 \catcode `\}=12
\catcode`\\=12 
|gdef|@alignverbatim#1\end{align}[#1|end[align]]
|gdef|@salignverbatim#1\end{align*}[#1|end[align*]]

|gdef|@alignatverbatim#1\end{alignat}[#1|end[alignat]]
|gdef|@salignatverbatim#1\end{alignat*}[#1|end[alignat*]]

|gdef|@xalignatverbatim#1\end{xalignat}[#1|end[xalignat]]
|gdef|@sxalignatverbatim#1\end{xalignat*}[#1|end[xalignat*]]

|gdef|@gatherverbatim#1\end{gather}[#1|end[gather]]
|gdef|@sgatherverbatim#1\end{gather*}[#1|end[gather*]]

|gdef|@gatherverbatim#1\end{gather}[#1|end[gather]]
|gdef|@sgatherverbatim#1\end{gather*}[#1|end[gather*]]

|gdef|@multilineverbatim#1\end{multiline}[#1|end[multiline]]
|gdef|@smultilineverbatim#1\end{multiline*}[#1|end[multiline*]]

|gdef|@arraxverbatim#1\end{arrax}[#1|end[arrax]]
|gdef|@sarraxverbatim#1\end{arrax*}[#1|end[arrax*]]

|gdef|@tabulaxverbatim#1\end{tabulax}[#1|end[tabulax]]
|gdef|@stabulaxverbatim#1\end{tabulax*}[#1|end[tabulax*]]

|endgroup

\def\align{\@verbatim \frenchspacing\@vobeyspaces \@alignverbatim
You are using the "align" environment in a style in which it is not defined.}

\@namedef{align*}{\@verbatim\@salignverbatim
You are using the "align*" environment in a style in which it is not defined.}
\expandafter\let\csname endalign*\endcsname =\endtrivlist

\def\alignat{\@verbatim \frenchspacing\@vobeyspaces \@alignatverbatim
You are using the "alignat" environment in a style in which it is not defined.}

\@namedef{alignat*}{\@verbatim\@salignatverbatim
You are using the "alignat*" environment in a style in which it is not defined.}
\expandafter\let\csname endalignat*\endcsname =\endtrivlist

\def\xalignat{\@verbatim \frenchspacing\@vobeyspaces \@xalignatverbatim
You are using the "xalignat" environment in a style in which it is not defined.}

\@namedef{xalignat*}{\@verbatim\@sxalignatverbatim
You are using the "xalignat*" environment in a style in which it is not defined.}
\expandafter\let\csname endxalignat*\endcsname =\endtrivlist

\def\gather{\@verbatim \frenchspacing\@vobeyspaces \@gatherverbatim
You are using the "gather" environment in a style in which it is not defined.}

\@namedef{gather*}{\@verbatim\@sgatherverbatim
You are using the "gather*" environment in a style in which it is not defined.}
\expandafter\let\csname endgather*\endcsname =\endtrivlist

\def\multiline{\@verbatim \frenchspacing\@vobeyspaces \@multilineverbatim
You are using the "multiline" environment in a style in which it is not defined.}

\@namedef{multiline*}{\@verbatim\@smultilineverbatim
You are using the "multiline*" environment in a style in which it is not defined.}
\expandafter\let\csname endmultiline*\endcsname =\endtrivlist

\def\arrax{\@verbatim \frenchspacing\@vobeyspaces \@arraxverbatim
You are using a type of "array" construct that is only allowed in AmS-LaTeX.}

\def\tabulax{\@verbatim \frenchspacing\@vobeyspaces \@tabulaxverbatim
You are using a type of "tabular" construct that is only allowed in AmS-LaTeX.}

\@namedef{arrax*}{\@verbatim\@sarraxverbatim
You are using a type of "array*" construct that is only allowed in AmS-LaTeX.}
\expandafter\let\csname endarrax*\endcsname =\endtrivlist

\@namedef{tabulax*}{\@verbatim\@stabulaxverbatim
You are using a type of "tabular*" construct that is only allowed in AmS-LaTeX.}
\expandafter\let\csname endtabulax*\endcsname =\endtrivlist

\def\@@eqncr{\let\@tempa\relax
    \ifcase\@eqcnt \def\@tempa{& & &}\or \def\@tempa{& &}%
      \else \def\@tempa{&}\fi
     \@tempa
     \if@eqnsw
        \iftag@
           \@taggnum
        \else
           \@eqnnum\stepcounter{equation}%
        \fi
     \fi
     \global\tag@false
     \global\@eqnswtrue
     \global\@eqcnt\z@\cr}

 \def\endequation{%
     \ifmmode\ifinner 
      \iftag@
        \addtocounter{equation}{-1} 
        $\hfil
           \displaywidth\linewidth\@taggnum\egroup \endtrivlist
        \global\tag@false
        \global\@ignoretrue   
      \else
        $\hfil
           \displaywidth\linewidth\@eqnnum\egroup \endtrivlist
        \global\tag@false
        \global\@ignoretrue 
      \fi
     \else   
      \iftag@
        \addtocounter{equation}{-1} 
        \eqno \hbox{\@taggnum}
        \global\tag@false%
        $$\global\@ignoretrue
      \else
        \eqno \hbox{\@eqnnum}
        $$\global\@ignoretrue
      \fi
     \fi\fi
 } 

 \newif\iftag@ \tag@false
 
 \def\tag{\@ifnextchar*{\@tagstar}{\@tag}}
 \def\@tag#1{%
     \global\tag@true
     \global\def\@taggnum{(#1)}}
 \def\@tagstar*#1{%
     \global\tag@true
     \global\def\@taggnum{#1}%
}

\makeatother

\begin{document}

\author{G. Lopez Castro$^{1}$, J. Pestieau$^{2}$, C. Smith$^{2}$ and S. Trine$^{2}$.
\and \quad \and \quad \\
$^{1}${\it \ Departamento F\'{i}sica, Centro de Investigaci\'{o}n y de
Estudios}\\
{\it \ Avanzados del IPN, Apdo. Postal 14-740, 07000 M\'{e}xico, D.F.,
M\'{e}xico}\\
$^{2}${\it \ Institut de Physique Th\'{e}orique, Universit\'{e} Catholique
de }\\
{\it Louvain, Chemin du Cyclotron 2, B-1348 Louvain-la-Neuve, Belgium}\quad 
\\
\quad \\
\quad \\
\quad}
\title{{\huge Parapositronium Decay and Dispersion Relations}\\
\quad \\
\quad \\
\quad \\
}
\maketitle

\begin{abstract}
\quad \newline
\quad \newline
Positronium decay rates are computed at the one-loop level, using
convolution-type factorized amplitudes. The dynamics of this factorization
is probed with dispersion relations, showing that unallowed approximations
are usually made, and some ${\cal O}\left( \alpha ^{2}\right) $ corrections
missed. Further, we discuss the relevance of the Schr\"{o}dinger
wavefunction as the basis for perturbative calculations. Finally, we apply
our formalism to the parapositronium two-photon decay.

\quad \newline

\quad \newline

\quad \newline

\quad \newline

\quad \newline

PACS Nos : 36.10.Dr, 12.20.Ds, 11.10.St, 11.55.Fv
\end{abstract}

\pagebreak

\section{Introduction}

Positronium is a bound state of electron and positron. In this paper, we
will be interested in the singlet state, parapositronium, whose decay rate
into $2\gamma $ has been precisely measured \cite{Experm}: 
\begin{equation}
\Gamma ^{exp}\left( p\text{-}Ps\rightarrow 2\gamma \right) =\left( 7.9909\pm
0.0017\right) \times 10^{9}\sec ^{-1}  \label{ExpValue}
\end{equation}

The corresponding theoretical predictions which include perturbative QED
corrections to a non-relativistic treatment of the bound state wavefunction
have been computed also with high accuracy (see for example \cite{KnownCorr}%
, \cite{Adkins}, \cite{BS2}): 
\begin{eqnarray*}
\Gamma \left( p\text{-}Ps\rightarrow 2\gamma \right) &=&\frac{\alpha ^{5}m}{2%
}\left[ 1-\left( 5-\frac{\pi ^{2}}{4}\right) \frac{\alpha }{\pi }+2\alpha
^{2}\log \frac{1}{\alpha }+1.75\left( 30\right) \left( \frac{\alpha }{\pi }%
\right) ^{2}-\frac{3\alpha ^{3}}{2\pi }\log ^{2}\frac{1}{\alpha }\right] \\
&=&7.98950\left( 2\right) \times 10^{9}\sec ^{-1}
\end{eqnarray*}
with $\alpha $ the fine structure constant and $m$ the electron mass. As can
be observed, agreement between theory and experiment is good. However, the
non-logarithmic ${\cal O}\left( \alpha ^{2}\right) $ corrections, which have
been obtained only recently, are not yet accessible experimentally.

Positronium is a test ground for bound state treatment in Quantum Field
Theory. The first try dates back to the $40^{\prime }s$, with decay rates
expressed through a factorized formula \cite{History} 
\[
\Gamma \left( p\text{-}Ps\rightarrow 2\gamma \right) =\left| \phi
_{o}\right| ^{2}\cdot \left( 4v_{rel}\sigma \left( e^{+}e^{-}\rightarrow
2\gamma \right) \right) _{v_{rel}\rightarrow 0} 
\]
with $\phi _{o}$ the Schr\"{o}dinger positronium wavefunction at the origin, 
$\sigma \left( e^{+}e^{-}\rightarrow 2\gamma \right) $ the total cross
section for $e^{+}e^{-}\rightarrow 2\gamma $ and $v_{rel}$ the relative
velocity of $e^{+}$ and $e^{-}$ in their center of mass frame. Since then,
more sophisticated decay amplitudes have been constructed, and systematic
procedures for calculating corrections have been developed. However, the
basic factorization of the bound state dynamics from the annihilation
process has remained as a basic postulate. For low order corrections, this
approximation is unquestionable, but for ${\cal O}\left( \alpha ^{2}\right) $
corrections, factorization has to be tested. Indeed, non-perturbative
phenomena responsible for the off-shellness of the electron and positron
inside the positronium are of ${\cal O}\left( \alpha ^{2}\right) $. In other
words, to get a sensible theoretical prediction at ${\cal O}\left( \alpha
^{2}\right) $, one must carefully analyze how binding energy effects enter
the general factorization approach.

In the present paper, we propose a systematic procedure for factorizing the
bound state dynamics from the annihilation process. From a fully
relativistic model at the one-loop level, where off-shell constituents
appear, we will recover the standard factorized amplitude used in the
literature. Most importantly, we will show that those standard formulas
involve some unnecessary approximations, and we will remove them. Since our
derivation relies on well-established techniques of quantum field theory, we
conclude that some ${\cal O}\left( \alpha ^{2}\right) $ corrections have
been forgotten. Our derivation of the factorized formula is particularized
to the parapositronium two-photon decay for the sake of definiteness, but it
is completely general, so equally valid for orthopositronium decays.

After those general considerations, an alternative and equivalent factorized
amplitude is found for parapositronium, which allows some further analyses
of the factorized formula. Using this last form, we compute lowest order
plus binding energy corrections for parapositronium decay. Those
calculations are carried with two different input forms for the bound state
wavefunction; first the well-known Schr\"{o}dinger momentum wavefunction,
and then an improved form for this wavefunction. Finally, we comment on
those two forms, and point towards the fact that it is possibly not the
usual simple Schr\"{o}dinger wavefunction, but rather the improved one that
should be used, due to the dynamics of the factorization.

\section{The Loop Model for Positronium Decay}

In this section, we introduce the loop model we will use to describe
positronium decay.

In that model, the positronium decays into a virtual electron-positron pair
which subsequently annihilates into real or virtual photons (an odd number
for ortho-states, an even number for para-states). The coupling of the
positronium to its constituents is described by a form factor, denoted by $%
F_{B}$. It is not assumed to be a constant, since a constant form factor
would amount to consider positronium as a point-like bound state. Specific
forms for $F_{B}$ will be discussed later; for now we just mention that it
should somehow be related to the bound state wavefunction.

For parapositronium decay into two photons, our model is represented by
figure 1. The corresponding amplitude is written 
\begin{equation}
{\cal M}^{\mu \nu }\left( p\text{-}Ps\rightarrow 2\gamma \right) =\int \frac{%
d^{4}q}{\left( 2\pi \right) ^{4}}F_{B}Tr\left\{ \gamma _{5}\frac{i}{%
\!\not\!%
q-\frac{1}{2}%
\!\not\!%
P-m}\Gamma ^{\mu \nu }\frac{i}{%
\!\not\!%
q+\frac{1}{2}%
\!\not\!%
P-m}\right\}  \label{ParaDec}
\end{equation}
with $m$ the electron mass and $F_{B}\equiv F_{B}\left( q^{2},P\cdot
q\right) $. The tensor $\Gamma ^{\mu \nu }$ is the scattering amplitude for
off-shell $e^{+}e^{-}$, with incoming momenta $\frac{1}{2}P-q$ and $\frac{1}{%
2}P+q$, into two photons : 
\[
\Gamma ^{\mu \nu }\left( e^{+}e^{-}\rightarrow \gamma \gamma \right)
=ie\gamma ^{\mu }\frac{i}{%
\!\not\!%
q-\frac{1}{2}%
\!\not\!%
P+%
\!\not\!%
l_{1}-m}ie\gamma ^{\nu }+ie\gamma ^{\nu }\frac{i}{%
\!\not\!%
q+\frac{1}{2}%
\!\not\!%
P-%
\!\not\!%
l_{1}-m}ie\gamma ^{\mu } 
\]
The decay width is expressed in terms of this amplitude 
\[
\Gamma \left( p\text{-}Ps\rightarrow 2\gamma \right) =\frac{1}{2!}\frac{1}{2M%
}\int d\Phi _{\gamma \gamma }\sum_{pol}\left| {\cal M}^{\mu \nu }\left( p%
\text{-}Ps\rightarrow 2\gamma \right) \varepsilon _{1\mu }^{*}\varepsilon
_{2\nu }^{*}\right| ^{2} 
\]
where $M<2m$ is the positronium mass.

This model is readily extended to any final states by changing the
scattering amplitude $\Gamma $ to the proper one, and to orthopositronium
states by the replacement $\gamma _{5}\rightarrow 
\!\not\!%
e$ with $e$ the orthopositronium polarization vector. Therefore, the
conclusions reached in the next section will be of complete generality.

\section{The Standard Approach as a Dispersion Relation}

What we intend to show in this section is that the formula (\ref{ParaDec})
leads to the expression for the decay amplitude found in the literature (see
for example \cite{KnownCorr}, \cite{Adkins}, \cite{TextBook1}) 
\begin{equation}
{\cal M}\left( p\text{-}Ps\rightarrow 2\gamma \right) \sim \int \frac{d^{3}%
{\bf k}}{\left( 2\pi \right) ^{3}2E_{{\bf k}}}\psi \left( {\bf k}^{2}\right)
Tr\left\{ \left( 1+\gamma ^{0}\right) \gamma _{5}\Gamma ^{\mu \nu }\left(
k,k^{\prime },l_{1}\right) \right\} \varepsilon _{1\mu }^{*}\varepsilon
_{2\nu }^{*}  \label{LiteraAmp}
\end{equation}
if we define the form factor as $F_{B}\equiv C\psi \left( {\bf k}^{2}\right)
\left( {\bf k}^{2}+\gamma ^{2}\right) $; $\psi \left( {\bf k}\right) $ is
the bound state wavefunction and $\gamma ^{2}=m^{2}-M^{2}/4$ (related to the
binding energy through $E_{B}=M-2m\approx -m\alpha ^{2}/4$). The scattering
amplitude $\Gamma ^{\mu \nu }$ describes the process $e^{-}\left( k\right)
e^{+}\left( k^{\prime }\right) \rightarrow 2\gamma $ with on-shell electron
and positron of momenta ${\bf k}$ and ${\bf k}^{\prime }=-{\bf k}$ and
energies $E_{{\bf k}}=E_{{\bf k}^{\prime }}=\sqrt{{\bf k}^{2}+m^{2}}$.

An interesting feature of this formula is that energy is apparently not
conserved, since the electron and positron have energies $E_{{\bf k}}>M/2$.
The justification of this formula as a simple convolution between the
positronium wavefunction and the scattering amplitude for $%
e^{+}e^{-}\rightarrow 2\gamma $ is therefore inadequate. What we are going
to demonstrate now is that one should understand (\ref{LiteraAmp}) as a
dispersion integral along the loop model branch cut. Further, the appearance
of the spin wavefunction of the bound state $\left( 1+\gamma ^{0}\right)
\gamma _{5}$ is questionable. Indeed, it is clear that it is the {\it moving}
electron-positron pair which has to be projected onto the required spin
state, so the projector cannot be simply $\left( 1+\gamma ^{0}\right) \gamma
_{5}$. This is confirmed by the fact that (\ref{LiteraAmp}) will be found by
neglecting some momentum $\left| {\bf k}\right| $ dependence in the exact
formula (\ref{ParaDec}). Taking into account the exact projector will
introduce some forgotten corrections.

Let us now demonstrate that by using a dispersion relation (see \cite
{TextBook2}, \cite{Kniehl}) to express the loop integration of (\ref{ParaDec}%
), we will reach (\ref{LiteraAmp}). Let us emphasize once again that the
whole discussion of this section is readily extended to any para- or
orthopositronium decay amplitude.

We first compute the imaginary part of (\ref{ParaDec}), $\func{Im}{\cal T}%
_{fi}\left( P^{2}\right) \equiv \func{Im}{\cal M}^{\mu \nu }\left( p\text{-}%
Ps\rightarrow 2\gamma \right) \varepsilon _{1\mu }^{*}\varepsilon _{2\nu
}^{*}$, for an arbitrary initial mass $P^{2}$. Considering the two possible
cuts (figure 2), we obtain $\func{Im}{\cal T}_{fi}$ by replacing the two
propagators on each side of $\Gamma ^{\mu \nu }$ by delta functions 
\begin{eqnarray*}
\func{Im}{\cal T}_{fi}\left( P^{2}\right) &=&\int \frac{d^{4}q}{2\left( 2\pi
\right) ^{2}}F_{B}\delta \left( \left( q-\frac{P}{2}\right)
^{2}-m^{2}\right) \delta \left( \left( q+\frac{P}{2}\right) ^{2}-m^{2}\right)
\\
&&\quad \times Tr\left\{ \gamma _{5}\left( 
\!\not\!%
q-\frac{%
\!\not\!%
P}{2}+m\right) \Gamma ^{\mu \nu }\left( 
\!\not\!%
q+\frac{%
\!\not\!%
P}{2}+m\right) \right\} \varepsilon _{1\mu }^{*}\varepsilon _{2\nu }^{*}
\end{eqnarray*}
After a straightforward integration over $q^{0}$ and $\left| {\bf q}\right| $%
, with $P=\left( \sqrt{P^{2}},{\bf 0}\right) $, we reach 
\begin{eqnarray*}
\func{Im}{\cal T}_{fi}\left( P^{2}\right) &=&\frac{1}{16\pi }\sqrt{1-\frac{%
4m^{2}}{P^{2}}}\theta \left( P^{2}-4m^{2}\right) \int \frac{d\Omega _{{\bf q}%
}}{4\pi }F_{B} \\
&&\quad \times Tr\left\{ \gamma _{5}\left( 
\!\not\!%
q-\frac{%
\!\not\!%
P}{2}+m\right) \Gamma ^{\mu \nu }\left( 
\!\not\!%
q+\frac{%
\!\not\!%
P}{2}+m\right) \right\} \varepsilon _{1\mu }^{*}\varepsilon _{2\nu }^{*}
\end{eqnarray*}
In the course of the derivation, the delta functions forced $q^{0}=0$ and $%
\left| {\bf q}\right| =\sqrt{P^{2}/4-m^{2}}$. In other words, the electron
momenta are 
\begin{equation}
\frac{1}{2}P\pm q=\left( \sqrt{\frac{P^{2}}{4}},\pm {\bf q}\right) \text{
with }\left( \frac{1}{2}P\pm q\right) ^{2}=\frac{P^{2}}{4}-\left| {\bf q}%
\right| ^{2}=m^{2}  \label{KinCut}
\end{equation}
This kinematics is to be understood in the trace evaluation. The angular
dependence arises from the relative orientations of ${\bf q}$ and the photon
momenta ${\bf l}_{1}$. Note also that the relation (\ref{KinCut}) cannot be
satisfied for the physical value $P^{2}=M^{2}<4m^{2}$. This is obvious since
the loop cannot have an imaginary part for the physical bound states, its
constituents being always off-shell. From the kinematics (\ref{KinCut}) one
can prove that the factors on both sides of $\Gamma ^{\mu \nu }$ are true
projectors, which serves to enforce gauge invariance in the expression 
\[
\left( 
\!\not\!%
q-\frac{%
\!\not\!%
P}{2}+m\right) \Gamma ^{\mu \nu }\left( 
\!\not\!%
q+\frac{1}{2}%
\!\not\!%
P+m\right) 
\]
Indeed, those two projectors play exactly the same role as external spinors
when demonstrating Ward identities. Remark also that the dependence of the
form factor on the loop energy can be taken arbitrarily since $q^{0}=0$.

The real part will now be calculated using an unsubstracted dispersion
relation 
\begin{equation}
\func{Re}{\cal T}_{fi}\left( M^{2}\right) =\frac{P}{\pi }\int_{4m^{2}}^{+%
\infty }\frac{ds}{s-M^{2}}\func{Im}{\cal T}_{fi}\left( s=P^{2}\right)
\label{ReTfi}
\end{equation}
where it is understood that $P^{2}$ should be replaced by $s$ everywhere,
i.e. scalar products that will appear when evaluating the trace should be
expressed with the kinematics defined for an initial energy $s$. Since $%
M^{2}<4m^{2}$, the principal part can be omitted and ${\cal T}_{fi}\left(
M^{2}\right) =\func{Re}{\cal T}_{fi}\left( M^{2}\right) $. Now let us write
the form factor in the general form 
\begin{equation}
F_{B}\equiv C\phi _{o}{\cal F}\left( {\bf q}^{2}\right) \left( {\bf q}%
^{2}+\gamma ^{2}\right) =C\phi _{o}{\cal F}\left( s/4-m^{2}\right) \cdot
\left( s-M^{2}\right) /4  \label{PostulatFb}
\end{equation}
with $\gamma ^{2}\equiv m^{2}-M^{2}/4$ and $\phi _{o}$ the bound state
wavefunction at zero separation. Then Eq. (\ref{ReTfi}) can be written as 
\begin{eqnarray*}
{\cal T}_{fi}\left( M^{2}\right) &=&C\phi _{o}\int_{4m^{2}}^{+\infty }ds\int 
\frac{d\Omega _{{\bf q}}}{4\pi }{\cal F}\left( s/4-m^{2}\right) \frac{\sqrt{%
1-4m^{2}/s}}{64\pi ^{2}} \\
&&\quad \times Tr\left\{ \gamma _{5}\left( 
\!\not\!%
q-\frac{%
\!\not\!%
P}{2}+m\right) \Gamma ^{\mu \nu }\left( 
\!\not\!%
q+\frac{%
\!\not\!%
P}{2}+m\right) \right\} \varepsilon _{1\mu }^{*}\varepsilon _{2\nu }^{*}
\end{eqnarray*}

Let us transform the $s$ integral back into a $\left| {\bf q}\right| $
integral, keeping in mind the constraints obtained when extracting the
imaginary part. Using ${\bf q}^{2}=s/4-m^{2}$, $ds=8\left| {\bf q}\right|
d\left| {\bf q}\right| $, the decay amplitude dispersion integral is 
\[
{\cal T}_{fi}\left( M^{2}\right) =\frac{C}{2}\phi _{o}\int \frac{d^{3}{\bf q}%
}{\left( 2\pi \right) ^{3}}\frac{{\cal F}\left( {\bf q}^{2}\right) }{\sqrt{%
P^{2}\left( {\bf q}\right) }}Tr\left\{ \gamma _{5}\left( 
\!\not\!%
q-\frac{%
\!\not\!%
P\left( {\bf q}\right) }{2}+m\right) \Gamma ^{\mu \nu }\left( 
\!\not\!%
q+\frac{%
\!\not\!%
P\left( {\bf q}\right) }{2}+m\right) \right\} \varepsilon _{1\mu
}^{*}\varepsilon _{2\nu }^{*} 
\]
where, as the notation suggests, it is understood that any $P^{2}$ appearing
in the amplitude must be replaced by $4\left| {\bf q}\right| ^{2}+4m^{2}$.
In particular, $\sqrt{P^{2}\left( {\bf q}\right) }$ can be replaced by $2E_{%
{\bf q}}$ with $E_{{\bf q}}=\sqrt{\left| {\bf q}\right| ^{2}+m^{2}}$. This
amounts to consider the scattering amplitude with incoming on-shell
electron-positron having momenta $\left( \frac{1}{2}P\left( {\bf q}\right)
\pm q\right) ^{2}=m^{2}$ (since $q^{0}=0$). Note the fact that $E_{{\bf q}%
}>M/2$, apparently the energy is not conserved. This is not surprising since
the present formula is a dispersion integral, done along the cut where $%
P^{2}\left( {\bf q}\right) >4m^{2}$. Finally, in view of the kinematics, we
introduce $k=\frac{1}{2}P\left( {\bf q}\right) +q$ and $k^{\prime }=\frac{1}{%
2}P\left( {\bf q}\right) -q$ (hence $E_{k}=E_{k^{\prime }}=E_{{\bf q}}$ and $%
{\bf k}=-{\bf k}^{\prime }={\bf q}$) to write the amplitude simply as 
\begin{equation}
{\cal T}_{fi}\left( M^{2}\right) =\frac{C}{2}\int \frac{d^{3}{\bf k}}{\left(
2\pi \right) ^{3}2E_{{\bf k}}}\left[ \phi _{o}{\cal F}\left( {\bf k}%
^{2}\right) \right] Tr\left\{ \gamma _{5}\left( -%
\!\not\!%
k^{\prime }+m\right) \Gamma ^{\mu \nu }\left( k,k^{\prime },l_{1}\right)
\left( 
\!\not\!%
k+m\right) \right\} \varepsilon _{1\mu }^{*}\varepsilon _{2\nu }^{*}
\label{StdDec}
\end{equation}
where $\Gamma ^{\mu \nu }\left( k,k^{\prime },l_{1}\right) $ is the
amplitude for on-shell $e^{-}\left( k\right) e^{+}\left( k^{\prime }\right) $
scattering into $2\gamma $. Gauge invariance is present due to the two
projectors, well defined since $k^{2}=k^{\prime 2}=m^{2}$. Hereafter, the
equation (\ref{StdDec}) will be referred as the standard approach
expression. Indeed, we can recognize (\ref{StdDec}) as the standard decay
amplitude for bound states \cite{TextBook1} : 
\begin{equation}
{\cal M}\left( p\text{-}Ps\rightarrow 2\gamma \right) =\sqrt{2M}\int \frac{%
d^{3}{\bf k}}{\left( 2\pi \right) ^{3}}\psi \left( {\bf k}\right) \frac{1}{%
\sqrt{2E_{{\bf k}}}}\frac{1}{\sqrt{2E_{{\bf k}}}}{\cal M}\left( e^{-}\left( 
{\bf k}\right) ,e^{+}\left( -{\bf k}\right) \rightarrow 2\gamma \right)
\label{StdStd*}
\end{equation}
with the amplitude constrained to the required spin state given by 
\[
{\cal M}\left( e^{-}\left( {\bf k}\right) ,e^{+}\left( -{\bf k}\right)
\rightarrow 2\gamma \right) =\frac{1}{2\sqrt{2}m}Tr\left\{ \gamma _{5}\left(
-%
\!\not\!%
k^{\prime }+m\right) \Gamma ^{\mu \nu }\left( k,k^{\prime },l_{1}\right)
\left( 
\!\not\!%
k+m\right) \right\} \varepsilon _{1\mu }^{*}\varepsilon _{2\nu }^{*} 
\]
Matching the present expression for the amplitude (\ref{StdDec}) with the
expression (\ref{StdStd*}), we obtain the constant $C$ 
\begin{equation}
C=\sqrt{M}/m  \label{CMatch}
\end{equation}
and the form factor ${\cal F}$ is identified with the wavefunction as $\phi
_{o}{\cal F}\left( {\bf k}^{2}\right) =\psi \left( {\bf k}\right) $. This
means that the function ${\cal F}\left( {\bf k}^{2}\right) $ is normalized
to unity and behaves as a delta of the momentum in the limit of vanishing
binding energy : 
\begin{equation}
\int \frac{d^{3}{\bf k}}{\left( 2\pi \right) ^{3}}{\cal F}\left( {\bf k}%
^{2}\right) =1,\quad \,\,\stackunder{\gamma \rightarrow 0}{\lim }{\cal F}%
\left( {\bf k}^{2}\right) =\left( 2\pi \right) ^{3}\delta ^{\left( 3\right)
}\left( {\bf k}\right)  \label{DeltaLimit}
\end{equation}

From the expression (\ref{StdDec}), it is clear that (\ref{LiteraAmp}) is an
approximation. Indeed, neglecting ${\bf k}$ dependences in the two
projectors $\left( -%
\!\not\!%
k^{\prime }+m\right) $ and $\left( 
\!\not\!%
k+m\right) $, we reach (\ref{LiteraAmp}) from (\ref{StdDec}). The projectors
appearing in (\ref{StdDec}) for particles in motion will introduce some new
corrections to the positronium decay rate of the order of the binding energy 
$\gamma ^{2}$, i.e. $\alpha ^{2}$. Further, when generalizing to
orthopositronium decay amplitudes, the expression (\ref{StdDec}) is gauge
invariant while (\ref{LiteraAmp}) is not. The fact that for the
parapositronium decay into two photons (\ref{LiteraAmp}) is gauge invariant
is due to the peculiar feature of $\gamma ^{5}$ appearance in the trace (see
(\ref{AmpTwoPhotI}) below).

In conclusion, the equivalence between the loop model and the standard
expression (\ref{StdDec}) for positronium decay amplitudes opens new
possibilities for explicit computations. This is precisely what we are going
to exploit in the following sections.

\section{Form Factor Dispersion Relation}

We have established the correspondence between (\ref{ParaDec}) and (\ref
{StdDec}). Let us construct an alternative, but equivalent, dispersion
procedure specific to the two-photon case that will be used in explicit
calculations. By computing the trace in (\ref{ParaDec}), the tensor
structure factorizes 
\begin{equation}
{\cal M}\left( p\text{-}Ps\rightarrow \gamma \gamma \right)
=8me^{2}\varepsilon ^{\mu \nu \rho \sigma }l_{1,\rho }l_{2,\sigma
}\varepsilon _{1\mu }^{*}\varepsilon _{2\nu }^{*}{\cal I}\left( M^{2}\right)
\label{AmpTwoPhotI}
\end{equation}
In this equation, ${\cal I}\left( M^{2}\right) $ can be viewed as an
effective form factor, modeled as the electron-positron loop with the
coupling $F_{B}$. There is only one term in ${\cal I}\left( M^{2}\right) $
since the direct and crossed amplitudes are equal under $q\rightarrow -q,$
i.e. an allowed variable change as $F_{B}\left( q^{2},P\cdot q\right)
=F_{B}\left( q^{2},-P\cdot q\right) $, and we write 
\begin{equation}
{\cal I}\left( P^{2}\right) =\eta \int \frac{d^{4}q}{\left( 2\pi \right) ^{4}%
}F_{B}\frac{1}{\left( q-\frac{1}{2}P\right) ^{2}-m^{2}}\frac{1}{\left( q+%
\frac{1}{2}P\right) ^{2}-m^{2}}\frac{1}{\left( q-\frac{1}{2}P+l_{1}\right)
^{2}-m^{2}}  \label{TwoPhotInt}
\end{equation}
It is to evaluate the effective form factor ${\cal I}\left( P^{2}\right) $
that we will now use dispersion techniques. Remark that the factorization of
the tensor part is interesting, since gauge invariance is manifest, and that 
${\cal I}\left( P^{2}\right) $ is convergent while the amplitude (\ref
{ParaDec}) is superficially divergent.

The factor $\eta $ is introduced because there is a subtlety in the above
factorization. Indeed, there is an arbitrariness in the choice of variable
for the dispersion integral. This situation is well-known for the photon
vacuum polarization : 
\begin{equation}
\Pi ^{\mu \nu }\left( k^{2}\right) =\left( k^{2}g^{\mu \nu }-k^{\mu }k^{\nu
}\right) \Pi _{1}\left( k^{2}\right) =\left( g^{\mu \nu }-\frac{k^{\mu
}k^{\nu }}{k^{2}}\right) \Pi _{2}\left( k^{2}\right)  \label{PhotPol}
\end{equation}
where one writes a dispersion relation for $\Pi _{1}\left( k^{2}\right) $,
which is less divergent than $\Pi _{2}\left( k^{2}\right) $ due to the
factorization of the tensor structure. The analogue of (\ref{PhotPol}) here
is 
\[
{\cal M}\left( p\text{-}Ps\rightarrow \gamma \gamma \right)
=8me^{2}\varepsilon ^{\mu \nu \rho \sigma }l_{1,\rho }l_{2,\sigma
}\varepsilon _{1\mu }^{*}\varepsilon _{2\nu }^{*}{\cal I}_{1}\left(
M^{2}\right) =8me^{2}\varepsilon ^{\mu \nu \rho \sigma }\frac{l_{1,\rho }}{M}%
\frac{l_{2,\sigma }}{M}\varepsilon _{1\mu }^{*}\varepsilon _{2\nu }^{*}{\cal %
I}_{2}\left( M^{2}\right) 
\]
Here we will choose to write a dispersion relation for ${\cal I}_{2}\left(
M^{2}\right) $, and this corresponds to the choice $\eta =P^{2}/M^{2}$. This
choice seems arbitrary, but is in fact necessary to recover the results of
the first section, i.e. the standard expression (\ref{StdStd*}). To
understand this, consider the factorized amplitude (\ref{AmpTwoPhotI}). In
the dispersion procedure used to recover the standard approach in the
previous section, the dispersion relation was built on the whole amplitude,
so it is clear that the photon momenta appearing in the tensor structure
were also incorporated, i.e. they were reduced as $l_{1,\rho }\rightarrow 
\overline{l_{1,\rho }}\times \sqrt{P^{2}}/2$. That is the reason why we must
include a factor $P^{2}$ into the effective form factor ${\cal I}\left(
P^{2}\right) $.

Let us give a general expression for ${\cal I}\left( P^{2}\right) $ as a
dispersion integral. The loop integral ${\cal I}\left( P^{2}\right) $ has an
imaginary part obtained by cutting the propagators 
\begin{eqnarray}
\func{Im}{\cal I}\left( P^{2}\right) &=&\frac{1}{2}\frac{P^{2}}{M^{2}}\int 
\frac{d^{4}q}{\left( 2\pi \right) ^{4}}F_{B}\frac{2\pi i\delta \left( \left(
q-\frac{1}{2}P\right) ^{2}-m^{2}\right) 2\pi i\delta \left( \left( q+\frac{1%
}{2}P\right) ^{2}-m^{2}\right) }{\left( q-\frac{1}{2}P+l_{1}\right)
^{2}-m^{2}}  \nonumber \\
&=&\frac{F_{B}}{8\pi M^{2}}\int \frac{d\Omega _{{\bf q}}}{4\pi }\frac{\sqrt{%
1-4m^{2}/s}}{1+\sqrt{1-4m^{2}/s}\cos \theta }\theta \left( s-4m^{2}\right)
\label{ImInt}
\end{eqnarray}
with $s=P^{2}$. The form factor is evaluated for $\left| {\bf q}\right|
^{2}=s/4-m^{2}$ and $q^{0}=0$. The unsubstracted dispersion relation is 
\begin{equation}
\func{Re}{\cal I}\left( M^{2}\right) =\frac{P}{\pi }\int_{-\infty }^{+\infty
}\frac{ds}{s-M^{2}}\func{Im}{\cal I}\left( s\right)  \label{RealDispInt}
\end{equation}
In our domain $M^{2}<4m^{2}$ and ${\cal I}\left( M^{2}\right) =\func{Re}%
{\cal I}\left( M^{2}\right) $. With the change of variable $\left| {\bf q}%
\right| ^{2}=s/4-m^{2}$, $ds=8\left| {\bf q}\right| d\left| {\bf q}\right| $%
, we write 
\begin{equation}
{\cal I}\left( M^{2}\right) =\func{Re}{\cal I}\left( M^{2}\right) =\frac{%
C\phi _{o}}{2M^{2}}\int \frac{d^{3}{\bf q}}{\left( 2\pi \right) ^{3}}{\cal F}%
\left( {\bf q}^{2}\right) \frac{1}{\sqrt{\left| {\bf q}\right| ^{2}+m^{2}}%
+\left| {\bf q}\right| \cos \theta }  \label{IntVarq}
\end{equation}
where $F_{B}\equiv C\phi _{o}{\cal F}\left( {\bf q}^{2}\right) \left( {\bf q}%
^{2}+\gamma ^{2}\right) $ (see \ref{PostulatFb}).

The final expression of the amplitude is therefore 
\begin{equation}
{\cal M}\left( p\text{-}Ps\rightarrow \gamma \gamma \right) =8me^{2}\frac{%
C\phi _{o}}{2M^{2}}\left[ \varepsilon ^{\mu \nu \rho \sigma }l_{1,\rho
}l_{2,\sigma }\varepsilon _{1\mu }^{*}\varepsilon _{2\nu }^{*}\right] \int 
\frac{d^{3}{\bf q}}{\left( 2\pi \right) ^{3}}{\cal F}\left( {\bf q}%
^{2}\right) \frac{1}{\sqrt{\left| {\bf q}\right| ^{2}+m^{2}}+\left| {\bf q}%
\right| \cos \theta }  \label{DispFbFinal}
\end{equation}
and the decay rate is expressed as 
\begin{equation}
\Gamma \left( p\text{-}Ps\rightarrow \gamma \gamma \right) =\frac{4\pi
\alpha ^{2}m^{2}}{M}C^{2}\left| \phi _{o}\right| ^{2}\left| \int \frac{d^{3}%
{\bf q}}{\left( 2\pi \right) ^{3}}{\cal F}\left( {\bf q}^{2}\right) \frac{1}{%
\sqrt{\left| {\bf q}\right| ^{2}+m^{2}}+\left| {\bf q}\right| \cos \theta }%
\right| ^{2}  \label{WidthTwoPhot}
\end{equation}

This is our third representation for the same decay amplitude : the first is
the loop integral (\ref{ParaDec}), the second is the well-known amplitude (%
\ref{StdDec}) or (\ref{StdStd*}) (with no approximation for projectors)
viewed as a dispersion integral for the amplitude, and the third is the
present dispersion integral (\ref{DispFbFinal}) for the effective loop form
factor ${\cal I}\left( P^{2}\right) $. All three procedures are strictly
equivalent to each other.

\section{Positronium Decay to two Photons}

Before going through the rate calculation, we will analyze the delta limit
for the form factor as in (\ref{DeltaLimit}). Then we will go through two
different calculations of $\Gamma \left( p\text{-}Ps\rightarrow \gamma
\gamma \right) $, obtained for specific choices of $F_{B}$ (or equivalently, 
${\cal F}\left( {\bf q}^{2}\right) $).

\subsection{Decay Rate in the Static Limit}

To compute the decay rate in the limit $\gamma ^{2}\rightarrow 0$ for the
form factor, we do not need to specify $F_{B}$. We just need to know that 
\[
F_{B}=C\phi _{o}{\cal F}\left( {\bf q}^{2}\right) \left( {\bf q}^{2}+\gamma
^{2}\right) \stackrel{\gamma ^{2}\rightarrow 0+}{\rightarrow }C\phi
_{o}\left( 2\pi \right) ^{3}\delta ^{\left( 3\right) }\left( {\bf q}\right)
\left( {\bf q}^{2}+\gamma ^{2}\right) 
\]
By setting ${\cal F}\left( {\bf q}^{2}\right) =\left( 2\pi \right)
^{3}\delta ^{\left( 3\right) }\left( {\bf q}\right) $ in (\ref{TwoPhotInt}),
we must recover exactly the lowest order decay rate $\Gamma \left( p\text{-}%
Ps\rightarrow \gamma \gamma \right) =\frac{1}{2}\alpha ^{5}m$. Using (\ref
{WidthTwoPhot}), we get 
\[
\int \frac{d^{3}{\bf q}}{\left( 2\pi \right) ^{3}}\left( 2\pi \right)
^{3}\delta ^{\left( 3\right) }\left( {\bf q}\right) \frac{1}{\sqrt{\left| 
{\bf q}\right| ^{2}+m^{2}}+\left| {\bf q}\right| \cos \theta }=\frac{1}{m} 
\]
Importantly, this result is independent of the binding energy : the loop do
not introduce any corrections in the static limit. The decay rate in that
limit is therefore : 
\begin{equation}
\Gamma \left( p\text{-}Ps\rightarrow \gamma \gamma \right) =\frac{1}{2}%
\alpha ^{5}m\left( \frac{m^{2}}{M}C^{2}\right)  \label{WidthCMatch}
\end{equation}
with $\left| \phi _{0}\right| ^{2}=\alpha ^{3}m^{3}/8\pi $. It remains to
match $C$ such that purely kinematic corrections vanish ($M$ factors in the
above formula arise from products like $\left( l_{1}\cdot l_{2}\right)
=M^{2}/2$ and from the $1/2M$ decay width factor, while $m$ comes from
electron propagators in the loop and from the wavefunction $\phi _{0}$).
With the definition (\ref{CMatch}) $C=\sqrt{M}/m$, the decay rate is exactly 
$\Gamma \left( p\text{-}Dm\rightarrow \gamma \gamma \right) =\frac{1}{2}%
\alpha ^{5}m$ as it should. In other words, the value for $C$ obtained by
matching (\ref{StdDec}) and (\ref{StdStd*}) is such that no correction arise
from factor $M/m$ in the static limit.

To conclude, let us repeat that we have not specified the form factor. This
means that any form factor which has a three-dimensional delta function
limit for $\gamma ^{2}\rightarrow 0$ gives the correct lowest order decay
rate $\frac{1}{2}\alpha ^{5}m$. In the following, we shall present two
forms, both built on the Schr\"{o}dinger momentum wavefunction.

\subsection{Schr\"{o}dinger Form Factor}

We can now apply the formulas of the preceding section to write down the
dispersion integral for the form factor 
\begin{equation}
{\cal F}_{I}\left( {\bf q}^{2}\right) =\frac{8\pi \gamma }{\left( {\bf q}%
^{2}+\gamma ^{2}\right) ^{2}}  \label{SchroI}
\end{equation}
with $\gamma ^{2}$ related to the binding energy and the fine structure
constant through $E_{B}=M-2m=-m\alpha ^{2}/4$. This form factor is just the
Shr\"{o}dinger momentum wavefunction for the bound state (note that it
satisfies the properties (\ref{DeltaLimit}) as used in \cite{Adkins}).

By using the formula (\ref{ImInt}) and (\ref{PostulatFb}), the imaginary
part is after the angular integration 
\[
\func{Im}{\cal I}\left( s\right) =\frac{C\phi _{o}}{16\pi M^{2}}\times
\left[ \frac{8\pi \gamma }{\left( {\bf q}^{2}+\gamma ^{2}\right) ^{2}}\left( 
{\bf q}^{2}+\gamma ^{2}\right) \right] _{{\bf q}^{2}=s/4-m^{2}}\times \ln
\left[ \frac{1+\sqrt{1-\frac{4m^{2}}{s}}}{1-\sqrt{1-\frac{4m^{2}}{s}}}%
\right] \times \theta \left( s-4m^{2}\right) 
\]
We now integrate the imaginary part using the unsubstracted dispersion
relation (\ref{RealDispInt}). As always, $M^{2}<4m^{2}$ so that the
principal part can be forgotten and ${\cal I}\left( M^{2}\right) =\func{Re}%
{\cal I}\left( M^{2}\right) $. For the given form factor ${\cal F}_{I}\left( 
{\bf q}^{2}\right) $, the calculation of integral ${\cal I}\left(
M^{2}\right) $ is now straightforward, and we get 
\begin{equation}
{\cal I}\left( M^{2}\right) =\frac{C\phi _{o}}{M^{3}}\frac{2}{\pi }\arctan 
\frac{M}{2\gamma }  \label{ReIntSchro1}
\end{equation}
The integral needed for this calculation is 
\begin{equation}
\int_{0}^{1}\frac{dx}{x_{o}-x}\,\,\ln \left[ \frac{1+\sqrt{1-x}}{1-\sqrt{1-x}%
}\right] \stackrel{x_{o}>1}{=}2\arctan ^{2}\frac{1}{\sqrt{x_{o}-1}}
\label{DispDJ}
\end{equation}
and its derivatives $(\partial \,/\partial x_{o})^{n}$. With the result for $%
{\cal I}$, the decay rate is 
\[
\Gamma \left( p\text{-}Ps\rightarrow \gamma \gamma \right) =\frac{1}{2}%
\alpha ^{5}m\left( \frac{4m^{2}}{M^{2}}\right) \left( \frac{2}{\pi }\arctan 
\frac{M}{2\gamma }\right) ^{2} 
\]
where we used $\left| \phi _{0}\right| ^{2}=\alpha ^{3}m^{3}/8\pi $ and $C=%
\sqrt{M}/m$. By expanding this result around $\gamma =0$, and expressing
corrections as a series in the fine structure constant $\alpha $, we recover
the standard result as zeroth order : 
\begin{eqnarray*}
\Gamma \left( p\text{-}Ps\rightarrow \gamma \gamma \right) &=&\frac{1}{2}%
\alpha ^{5}m\left( 1-\frac{\alpha }{\pi }+\frac{1}{8}\alpha ^{2}-\frac{13}{%
96\pi }\alpha ^{3}+\frac{1}{64}\alpha ^{4}+{\cal O}\left( \alpha ^{5}\right)
\right) ^{2} \\
&\approx &\frac{1}{2}\alpha ^{5}m\left( 1-0.637\alpha +0.351\alpha
^{2}-0.166\alpha ^{3}+0.074\alpha ^{4}+{\cal O}\left( \alpha ^{5}\right)
\right)
\end{eqnarray*}
Numerically, the corrections at the one-loop level are 
\[
\Gamma ^{Form\,Fact.}\left( p\text{-}Ps\rightarrow \gamma \gamma \right)
=\Gamma _{o}\times 0.9954\approx 7.9956\times 10^{9}\sec ^{-1} 
\]
If we combine the present correction with radiative corrections up to order $%
\alpha ^{2}\ln \alpha $ as found in the literature \cite{KnownCorr}, the
theoretical value is modified as 
\[
\Gamma ^{Rad.Corr.\,+\,Form\,Fact.}\left( p\text{-}Ps\rightarrow \gamma
\gamma \right) =7.9527\times 10^{9}\sec ^{-1} 
\]
while the experimental value is (\ref{ExpValue}). We can therefore note that
the present form factor leads to a too small value for $\Gamma \left( p\text{%
-}Ps\rightarrow \gamma \gamma \right) $, since it introduces new order $%
\alpha $ corrections. The problem is in the form factor, which do not
converge fast enough towards the static limit delta. In other words, for a
given $\gamma ^{2}$, ${\cal F}_{I}\left( {\bf q}^{2}\right) $ is not enough
peaked around ${\bf q}=0$.

\subsection{Improved Schr\"{o}dinger form factor}

The second possibility we analyze is 
\begin{equation}
{\cal F}_{II}\left( {\bf q}^{2}\right) =\frac{32\pi \gamma ^{3}}{\left( {\bf %
q}^{2}+\gamma ^{2}\right) ^{3}}  \label{SchroII}
\end{equation}
This form factor has the same delta limit property (\ref{DeltaLimit}), but
converges faster than (\ref{SchroI}). We can therefore expect much smaller
corrections than with ${\cal F}_{I}\left( {\bf q}^{2}\right) $. Repeating
the derivation of the preceding section, we get from (\ref{ImInt}) : 
\[
\func{Im}{\cal I}\left( s\right) =\frac{C\phi _{o}}{16\pi M^{2}}\times
\left[ \frac{32\pi \gamma ^{3}}{\left( {\bf q}^{2}+\gamma ^{2}\right) ^{3}}%
\left( {\bf q}^{2}+\gamma ^{2}\right) \right] _{{\bf q}^{2}=s/4-m^{2}}\times
\ln \left[ \frac{1+\sqrt{1-\frac{4m^{2}}{s}}}{1-\sqrt{1-\frac{4m^{2}}{s}}}%
\right] \times \theta \left( s-4m^{2}\right) 
\]
The unsubstracted dispersion integral is (\ref{RealDispInt}), and we get
using the result (\ref{DispDJ}) : 
\[
{\cal I}\left( M^{2}\right) =\func{Re}{\cal I}\left( M^{2}\right) =\frac{%
C\phi _{o}}{M^{3}}\left( \frac{32\gamma ^{3}}{\pi M^{3}}\right) \left( \frac{%
M^{2}}{8\gamma ^{2}}-\frac{M}{4\gamma }\left( 1-\frac{M^{2}}{4\gamma ^{2}}%
\right) \arctan \frac{M}{2\gamma }\right) 
\]
The rate is then 
\[
\Gamma \left( p\text{-}Ps\rightarrow \gamma \gamma \right) =\frac{1}{2}%
\alpha ^{5}m\left( \frac{4m^{2}}{M^{2}}\right) \left[ \left( \frac{32\gamma
^{3}}{\pi M^{3}}\right) \left( \frac{M^{2}}{8\gamma ^{2}}-\frac{M}{4\gamma }%
\left( 1-\frac{M^{2}}{4\gamma ^{2}}\right) \arctan \frac{M}{2\gamma }\right)
\right] ^{2} 
\]
Thus, transcribing into a series in $\alpha $ : 
\[
\Gamma \left( p\text{-}Ps\rightarrow \gamma \gamma \right) =\frac{1}{2}%
\alpha ^{5}m\left( 1-\frac{1}{4}\alpha ^{2}+\frac{2}{3\pi }\alpha ^{3}-\frac{%
7}{64}\alpha ^{4}+{\cal O}\left( \alpha ^{5}\right) \right) 
\]
Numerically, the corrections starting at order $\alpha ^{2}$ are 
\[
\Gamma \left( p\text{-}Ps\rightarrow \gamma \gamma \right) =\left(
1-1.32\times 10^{-5}\right) \times \Gamma _{o} 
\]
i.e. very small. If we combine the present correction with radiative
corrections up to order $\alpha ^{2}\ln \alpha $ as found in the literature,
the theoretical value is modified as 
\[
\Gamma ^{rad.Corr.\,+\,Form\,Fact.}\left( p\text{-}Ps\rightarrow \gamma
\gamma \right) =7.9894\times 10^{9}\sec ^{-1} 
\]
As announced, this form factor leads to acceptable corrections, contrary to
the ${\cal F}_{I}\left( {\bf q}^{2}\right) $ form factor.

Before closing this section, let us discuss how the form factor ${\cal F}%
_{II}\left( {\bf q}^{2}\right) $ could arise as the right form factor.
First, one can see that it implies the $F_{B}$ form 
\[
F_{B}=C\phi _{o}\frac{32\pi \gamma ^{3}}{\left( {\bf q}^{2}+\gamma
^{2}\right) ^{3}}\left( {\bf q}^{2}+\gamma ^{2}\right) =C\phi _{o}\left[ 
\frac{8\pi \gamma }{\left( {\bf q}^{2}+\gamma ^{2}\right) ^{2}}\right]
4\gamma ^{2} 
\]
The factor in brackets is the momentum Schr\"{o}dinger wavefunction while
the additional factor $\gamma ^{2}$ gives to the coupling $F_{B}$ the
desirable property of vanishing when $\gamma \rightarrow 0$. From known
Bethe-Salpeter analyses (see for example \cite{Adkins},\cite{BarbRem},\cite
{BS2}), one constructs an approximated bound state wavefunction by
considering only coulombic photon exchanges in the Bethe-Salpeter kernel.
Let $\Psi \left( q\right) $ be this wavefunction (the Barbieri-Remiddi
wavefunction \cite{BarbRem}), but with its spin wavefunction part omitted.
Then, if one defines the coupling as 
\begin{equation}
F_{B}=C\times 4\gamma ^{2}\times \left( {\bf q}^{2}+\gamma ^{2}\right)
\times \Psi \left( q\right)  \label{BStoFb}
\end{equation}
by going through the dispersion analyses of section 3, one ends up with $%
\psi \left( {\bf q}^{2}\right) ={\cal F}_{II}\left( {\bf q}^{2}\right) $.
The reason for this is quite technical, but let us just mention that usually
one ''uses'' some part of $\Psi \left( q\right) $ to set the energy $q^{0}$
to zero to go from a four-dimensional towards a three-dimensional
convolution-type decay amplitude (see \cite{Adkins}, \cite{BS2}), while here
it is the dispersion relation which enforces $q^{0}=0$, leaving $\Psi \left(
q\right) $ unaltered. This brings a supplementary $1/\left( {\bf q}%
^{2}+\gamma ^{2}\right) $ factor. Further, the formula (\ref{BStoFb}) is the
analogue of standard expressions making the connection between
Bethe-Salpeter vertex and bound state wavefunction via some propagators.
Even if the present remarks do not constitute a rigorous proof, they point
towards ${\cal F}_{II}\left( {\bf q}^{2}\right) $ as the appropriated form
factor, rather than the widely used ${\cal F}_{I}\left( {\bf q}^{2}\right) $%
. In conclusion, we would like to stress that the well-known
Barbieri-Remiddi wavefunction, used as a basis for decay rate calculations,
is not simply the Schr\"{o}dinger momentum wavefunction, contrary to what is
often seen in the literature.

\section{Conclusions}

The result of our analyses is three-fold. First, we have shown how the
standard convolution-type factorized amplitude can be given a coherent
grounding from dispersion relations. This demonstrates that the formula
usually quoted is an approximation, missing some of those ${\cal O}\left(
\alpha ^{2}\right) $ corrections it is meant to evaluate. Then we have
calculated those corrections for $p$-$Ps\rightarrow \gamma \gamma $ using
our exact factorized formula, or equivalently, its effective form factor
realization, for two different couplings $F_{B}$. Finally, we comment on
those couplings.

Concerning this last point, it should now be clear that the dispersion
method used to factorize the bound state dynamics from the annihilation
process imposes some constraints. By this we mean that there is a definite
procedure for gluing together the bound state wavefunction and the decay
amplitude.

More precisely, the energy dependences and the spinor part of the
Bethe-Salpeter wavefunction are usually incorrectly introduced. As we have
discussed, the energy dependences introduce a modification of the naive
Schr\"{o}dinger wavefunction, while the spinor part should be introduced in
the decay process rather that in the bound state wavefunction, in order to
properly project the constituents onto the required spin state.

Finally, we have shown that the static limit reproduces the first
approximation to decay rates. This approximation will be modified by higher
order corrections only, provided the form factor has a three-dimensional
delta limit for $\gamma \rightarrow 0$. Whether this limit property is
shared by quarkonia wavefunctions is still an open question.

\qquad \newline

In a following paper, the present formalism will be applied to the
orthopositronium decay into $\gamma \gamma \gamma $. The analytical defects
of this decay channel, noticed in \cite{AnsatzWork}, will be completely
removed thanks to dispersion techniques. As a result, in addition to the new 
${\cal O}\left( \alpha ^{2}\right) $ corrections found here, a whole set of
new amplitudes contributing to that order will be presented. Those
amplitudes are non-factorizable contributions to the decay rate that could
not appear in the standard factorized procedures, but unavoidable from the
requirements of gauge invariance and analyticity.

\qquad \newline

{\Large Acknowledgements: }G. L. C. was partially supported by Conacyt
(M\'{e}xico) under contract No 32429. C. S. and S. T. acknowledge financial
supports from FNRS (Belgium).


\begin{thebibliography}{99}
\bibitem{Experm}  A. Al-Ramadhan, D. Gidley, Phys. Rev. Lett. {\bf 72}, 1632
(1994).

\bibitem{KnownCorr}  I. Harris and L. Brown, Phys. Rev. {\bf 105}, 1656
(1957); W. E. Caswell, G. P. Lepage, Phys. Rev. {\bf A20}, 36 (1979); A.
Czarneski, K. Melnikov, A. Yelkhovsky, Phys. Rev. Lett. {\bf 83}, 1135
(1999); eprint {\it hep-ph/9910488} and references cited therein.

\bibitem{Adkins}  G. Adkins, Ann. Phys. {\bf 146}, 78 (1983).

\bibitem{BS2}  V. Antonelli, V. Ivanchenko, E. Kuraev, V. Laliena, Eur.
Phys. J. {\bf C5}, 535 (1998), V. Antonelli,{\it \ Int. Work. on Hadronic
Atoms and Positronium in the S.M.}, Dubna, 26-31 May 1998.

\bibitem{History}  J. A. Wheeler, Ann. N. Y. Acad. Sci. {\bf 48}, 219
(1946); J. Pirenne, Arch. Sci. Phys. Nat. {\bf 29}, 265 (1947).

\bibitem{TextBook1}  M. Peskin, D. Schroeder, {\it An Introduction to
Quantum Field Theory}, Addison-Wesley (1995); O. Nachtmann, {\it Elementary
Particle Physics : Concepts and Phenomena, }Springer-Verlag (1990)

\bibitem{BarbRem}  R. Barbieri and E. Remiddi, Nucl. Phys. {\bf B141}, 413
(1978).

\bibitem{TextBook2}  K. Nishijima, {\it Fields and Particles : Field Theory
and Dispersion Relations}, Benjamin N.Y. (1969).

\bibitem{Kniehl}  B. Kniehl, Acta Phys. Polon. {\bf B27}, 3631 (1996) ({\it %
hep-ph/9607255}).

\bibitem{AnsatzWork}  G. Lopez Castro, J. Pestieau and C. Smith, {\it %
hep-ph/0004209}.
\end{thebibliography}
\end{document}